\documentclass[hidelinks,onefignum,onetabnum]{siamart250211}

\usepackage{caption}
\usepackage{lipsum}
\usepackage{amsfonts}
\usepackage{graphicx}
\usepackage{epstopdf}
\usepackage{geometry}
\usepackage{amsmath,amssymb}
\usepackage{algorithm}
\usepackage{algorithmic}
\usepackage{bm}
\usepackage{appendix}
\usepackage{multirow}
\usepackage{braket}
\usepackage[english]{babel}
\usepackage{hyperref}
\usepackage[capitalize]{cleveref}
\usepackage[square,numbers,sort&compress]{natbib}
\usepackage{soul}
\usepackage{tikz}
\usetikzlibrary{quantikz2}
\usepackage{adjustbox}
\usepackage{xspace}
\ifpdf
  \DeclareGraphicsExtensions{.eps,.pdf,.png,.jpg}
\else
  \DeclareGraphicsExtensions{.eps}
\fi

\def \Im{\mathop{\rm Im}}

\newcommand{\REV}[1]{#1}

\numberwithin{equation}{section}

\newsiamremark{remark}{Remark}
\newsiamremark{hypothesis}{Hypothesis}
\crefname{hypothesis}{Hypothesis}{Hypotheses}
\newsiamthm{claim}{Claim}
\newsiamthm{example}{Example}

\headers{Qubit-Efficient Quantum Algorithm for Linear Differential Equations}{Di Fang, David Lloyd George and Yu Tong}

\title{Qubit-Efficient Quantum Algorithm for Linear Differential Equations
\thanks{
\funding{This work is supported by the U.S. Department of Energy, Office of Science, Accelerated Research in Quantum Computing Centers, Quantum Utility through Advanced Computational Quantum Algorithms, grant no. DE-SC0025572, and National Science Foundation via the grant DMS-2347791 and DMS-2438074.}}}

\author{Di Fang\thanks{Department of Mathematics and Duke Quantum Center, Duke University 
  (\email{di.fang@duke.edu}).}
\and David Lloyd George \thanks{Department of Physics and Duke Quantum Center, Duke University
  (\email{david.lloydgeorge@duke.edu}).}
  \and Yu Tong \thanks{Department of Mathematics, Department of Electrical and Computer Engineering and Duke Quantum Center, Duke University
  (\email{yu.tong@duke.edu}).}
  }

\usepackage{amsopn}

\usepackage{url}
\usepackage{enumitem}
\usepackage{indentfirst}

\numberwithin{equation}{section}

\renewcommand{\Re}{\operatorname{Re}}
\renewcommand{\Im}{\operatorname{Im}}

\newcommand{\norm}[1]{\left\lVert#1\right\rVert}

\newcommand{\Or}{\mathcal{O}}

\newcommand{\rd}{\mathrm{d}}

\ifpdf
\hypersetup{
  pdftitle={Qubit-Efficient Quantum Algorithm for Linear Differential Equations},
  pdfauthor={Di Fang, David Lloyd George, and Yu Tong}
}
\fi

\begin{document}

\maketitle

\begin{abstract}
As quantum hardware rapidly advances toward the early fault-tolerant era, a key challenge is to develop quantum algorithms that are not only theoretically sound but also hardware-friendly on near-term devices. In this work, we propose a quantum algorithm for solving linear ordinary differential equations (ODEs) with a provable runtime guarantee. Our algorithm uses only a single ancilla qubit, and is locality preserving, i.e., when the coefficient matrix of the ODE is $k$-local, the algorithm only needs to implement the time evolution of $(k+1)$-local Hamiltonians. We also discuss the connection between our proposed algorithm and Lindbladian simulation. By applying our algorithm to the interacting Hatano-Nelson model, a widely studied non-Hermitian model with rich phenomenology, and numerically simulating it under realistic noise models, we demonstrate its practical feasibility on near-term quantum devices.
\end{abstract}



\section{Introduction}

As the development of quantum computing hardware continues to march towards fault tolerance, a central challenge is how to meaningfully integrate advanced quantum algorithm theory with the evolving capabilities of quantum hardware--especially for high-impact applications. Such a perspective has attracted significant attention in the context of Hamiltonian simulation, but remains relatively underexplored for broader classes of problems such as solving general differential equations. At the same time, emerging hardware features, such as mid-circuit measurement and other architectural innovations, are opening up new opportunities for algorithm design. In this context, an important and timely question is how to adapt advanced theoretical algorithms to better align with early fault-tolerant platforms and make them practical for near-term quantum devices.

Quantum algorithms can be generally classified into two broad categories: heuristic algorithms, such as variational quantum algorithms, and fault-tolerant algorithms with provable performance guarantees. Heuristic algorithms are often considered near-term feasible, but it is typically challenging to establish rigorous guarantees for them. In contrast, fault-tolerant algorithms are designed with theoretical guarantees in mind, often achieving excellent asymptotic scaling. However, their practical implementation is usually beyond the capability of near-term devices due to the reliance on advanced quantum techniques -- such as those requiring large numbers of ancilla qubits and complex control circuits -- which remain out of reach for current hardware.

There is, however, a notable exception: Trotterization for Hamiltonian simulation~\cite{Trotter1959,Suzuki1993,WiebeBerryHoyerSanders2010,ChildsSuTranEtAl2020,ChildsSu2019,SahinogluSomma2020,AnFangLin2021,SuHuangCampbell2021,ZhaoZhouShawEtAk2021,BornsWeilFang2022,GongZhouLi2023,LowSuTongTran2023,ZhaoZhouChilds2024,MoralesCostaPantaleoniBurgarthSandersBerry2025}, which approximates the full unitary evolution by decomposing it into a product of simpler unitaries, each generated by an individual component of the Hamiltonian. This is a fault-tolerant algorithm with provable performance, yet it is remarkably simple and avoids the use of sophisticated quantum subroutines. As a result, Trotterization can often be implemented even on today's quantum devices (see for example \cite{MethZhangEtAl2025simulating,KangEtAl2024seeking,Whitlow2023quantum,Seetharam2023digital}). This simplicity -- requiring minimal ancilla (none in this case) and no complex control units -- has made the first-order Trotterization one of the most widely used and the most impactful algorithms, even though its asymptotic scaling may be worse than more advanced approaches such as quantum signal processing (QSP)~\cite{LowChuang2017}, quantum singular value transformation (QSVT)~\cite{GilyenSuLowEtAl2019}, or truncated Taylor, Dyson, or Magnus series expansions~\cite{BerryChildsCleveEtAl2015,KieferovaSchererBerry2019,LowWiebe2019,BerryChildsSuEtAl2020,AnFangLin2022,FangLiuSarkar2025,BornsweilFangZhang2025}.

Unfortunately, the favorable properties of Trotterization rely heavily on the specific structure of the problem it targets -- namely, unitary evolution governed by Hermitian Hamiltonians.
When extended to more general linear differential equations, which can describe non-unitary dynamics, this structure no longer holds unless in special cases where the norm remains preserved by the dynamics~\cite{BabbushBerryKothariSommaWiebe2023}. In such settings, some form of block encoding is required to represent the non-unitary evolution at each time step. This introduces a success probability associated with each application. Over $L$ time steps, the overall success probability can decay exponentially in $L$, making the naive application of Trotter-like ideas ineffective~\cite{Berry2014}. To overcome this, existing approaches typically resort to more advanced tools such as the quantum linear systems algorithm (QLSA) combined with encoding history states~\cite{Berry2014,BerryChildsOstranderWang2017,ChildsLiu2019,Krovi2022,BerryCosta2024}, compression gadgets for the time-marching strategy~\cite{FangLinTong2023,LowWiebe2019,YangOnwuntaAn2025}, linear combinations of unitaries (LCU) in linear combinations of Hamiltonian simulation~\cite{AnLiuLin2023linear,AnChildsLinYing2024laplace,AnChildsLin2023quantum}, Lindbladian solvers in \cite{ShangGuoAnZhao2024} and quantum eigenvalue transformation in \cite{LowSu2024}, or use structures present in the problem \cite{wright2024noisy,lubasch2025quantum,BabbushBerryKothariSommaWiebe2023,Goldfriend2025implementation}. Compared to the QLSA-based methods, the non-QLSA-based algorithms can often achieve state-preparation costs that are independent of the desired precision. In particular, when combined with amplitude amplification, all post-QLSA techniques attain optimal state-preparation scaling that matches the known lower bounds~\cite{AnLiuWangZhao2022}.
These techniques can successfully address the vanishing success probability issue, restore provable guarantees and can achieve excellent asymptotic cost scaling, but at the expense of significantly increased circuit resource requirements (such as the number of ancilla qubits and the complexity in the control units), 
making them challenging to implement on
near-term or current quantum devices.

This raises the natural question:
\begin{center}
\textit{
Can we design a quantum algorithm that is both near-term (without relying on advanced fault-tolerant subroutines) and with provable guarantees for linear differential equations?
}\end{center} 
We answer this question affirmatively by proposing a quantum algorithm for solving linear differential equations following the design principle of minimizing the number of ancilla qubits and avoiding complex algorithmic subroutines. Specifically, our algorithm uses only \textit{one} additional ancilla qubit and \textit{no} controlled unitary. The algorithm does not rely on any advanced fault-tolerant building blocks such as LCU, QSVT, or QLSA, making it suitable for near-term implementation in the early fault-tolerant era. This design principle is shared by many previous works on early fault-tolerant quantum algorithms \cite{Campbell2021early,LinTong2019,Tong2022designing,DongLinTong2022ground,DingDongTongLin2023robust,DingLin2023even,DingLin2023simultaneous,DingChenLin2024single,WangFrancaEtAl2023quantum,KatabarwaGratseaEtAl2024early,WangFrancaEtAl2025efficient,WangMcArdleBerta2024qubit,McArdleGilyenBerta2022quantum,KissAzadEtAl2025early,FomichevHejaziEtAl2024initial}.
\REV{We also demonstrate its feasibility on near-term quantum devices by classically  simulating its performance using the Qiskit SDK \cite{JavadiEtAl2024quantum} under realistic noise models, as discussed in detail in~\cref{sec:appl_Hatano-Nelson}. }

Like Trotterization for Hamiltonian simulation,  our goal is not to achieve optimal asymptotic scaling, but rather to design an algorithm that is practical in the near term while still providing rigorous performance guarantees. This is, to our knowledge, the first quantum algorithm for differential equations in this category, that is, both near-term friendly (without using advanced fault-tolerant subroutines) and suitable for the fault-tolerant setting (with theoretical guarantees).

\section{The Algorithm}
\label{sec:algorithm}

In this paper, we focus on the initial value problem of the system of linear ordinary differential equations (ODEs) 
\begin{equation} \label{eq:ode_general}
    \frac{\rd}{\rd t} \ket{\psi(t)} = A(t) \ket{\psi(t)}, \quad \ket{\psi(0)} = \ket{\psi_0},
\end{equation}
where $t \in [0, T] \subset \mathbb{R}^+$ is the independent variable with $T$ as the final time, the vector $ \ket{\psi(t)} \in \mathbb{C}^N$ is the dependent variable, the coefficient matrix $A(t) \in \mathbb{C}^{N\times N}$ is a matrix-valued function in $t$, and $\ket{\psi_0} \in \mathbb{C}^N$ is the initial condition. We assume the differential equation has a well-posed solution, which can be implied by, e.g., $A(t)$ being piecewise continuous. Differential equations of this type arise naturally from quantum problems, such as transcorrelated methods in quantum chemistry \cite{FomichevHejaziEtAl2024initial,MottaGujaratiEtAl2020quantum,LeeThom2023studies} and imaginary time evolution \cite{MottaSunTanEtAl2020}.

The coefficient matrix in \cref{eq:ode_general} can be written into a sum of Hermitian and anti-Hermitian parts 
\[
A = \underbrace{\frac{A+A^\dag}{2}}_{: = V} +   \underbrace{\frac{A-A^\dag}{2}}_{: = iH} .
\]
Under the dissipative condition, all the eigenvalues of the Hermitian part are non-positive. In other words, the Hermitian part is negative semi-definite, which is then equivalent to there exists a matrix $L$ such that $V = -L^\dagger L$. More generally, we consider the following linear differential equation
\begin{equation}
\label{eq:LL_decompose}
\begin{aligned}
    \frac{\rd}{\rd t} \ket{\psi(t)} &= A(t)\ket{\psi(t)} := \left(-iH(t) - \sum_{j = 1}^J L_j^\dagger L_j \right)\ket{\psi(t)}, \\
    \ket{\psi(0)} &= \ket{\psi_0}.
\end{aligned}
\end{equation}
The reason to consider a set of $L_j$'s instead of one is because such a decomposition is more flexible, and can utilize features of $A(t)$ such as locality to make it more suitable for near-term implementation. 
\REV{As an example, when the Hermitian part $V$ consists of local terms, i.e., $V=\sum_{j=1}^J V_j$, where each $V_j$ acts only on $k$ qubits (or fermionic modes), then one can find real number $c_j$ for each $V_j$ such that $V_j-c_j$ is negative semi-definite. The operators $L_j$ can then be constructed as $L_j=\sqrt{c_j-V_j}$. This procedure can be done efficiently on a classical computer due to the locality structure. A caveat of this example is that we would be implementing the evolution under $A-\sum_j c_j$ rather than $A$, which can lead to an artificial decay of the solution norm that leads to unfavorable computational cost. One way to minimize this artificial decay is to parameterize each $L_j$ as a linear combination of low-weight Pauli operators (e.g., $k/2$) and then optimize the coefficients. This can be cast into a semi-definite programming problem, and has}
recently been studied for electronic structure problems and for the Sachdev–Ye–Kitaev model~\cite{KingLowEtAl2025quantum,LowKingBerryEtAl2025fast}.
\REV{It is also possible to utilize structures in $A$ to go beyond the local case. For example, in~\cref{sec:convection_diffusion} we discuss how this decomposition can be obtained for the convection-diffusion equation, whose generator is not a linear combination of local Pauli operators.}

The fact that the Hermitian part $V$ is negative semidefinite implies that the norm of the solution can never grow:
\begin{equation}
    \label{eq:norm_decay}
    \frac{\rd}{\rd t}\braket{\psi(t)|\psi(t)} = 2\braket{\psi(t)|V|\psi(t)}\leq 0.
\end{equation}

We will Trotterize the time evolution so that we only need to implement the evolution governed by the anti-Hermitian part $-iH$ and the Hermitian (dissipative) part $-L^\dagger_j L_j$ separately for time $\tau$.
For the anti-Hermitian part, its evolution is described by the time-evolution operator $e^{-iH\tau}$ and can be implemented via standard Hamiltonian simulation. For the dissipative part, we consider the following Hermitian operator
\begin{equation}
\label{eq:defn_Gj}
G_j = \begin{pmatrix}
    0 & L_j^\dagger \\
    L_j & 0
\end{pmatrix},
\end{equation}
and one can readily check that
\begin{equation}
\label{eq:first_order_approx_dissipative_evolution}
     \left( \bra{0} \otimes I \right) 
    e^{ i\sqrt{2\tau}  G_j} \ket{0} \ket{\psi}
=
\left( I - \tau L_j^\dagger L_j \right) \ket{\psi} + \Or(\tau^2),
\end{equation}
which implements a first-order approximation of $e^{-\tau L_j^\dagger L_j}$ acting on the state $\ket{\psi}$. In particular, it is convenient to observe that the odd-order terms with respect to $\sqrt{\tau}$ in the Taylor expansion vanish because $G_j$ is an off-diagonal block matrix, and in its block matrix representation, the state $\ket{0} \ket{\psi}$ has a second block equal to zero. Note that the operator $e^{ i\sqrt{2\tau}  G_j}$ is the time evolution operator generated by a Hermitian Hamiltonian $G_j$, and can therefore be implemented through Hamiltonian simulation. This observation forms the basis of our algorithm.

The algorithm can be described as follows:

\begin{itemize}
    \item \textbf{Initialization:} We start with the state $\ket{0} \ket{\psi_0}$, where the first register serves as a purification register containing one ancilla qubit, and the second register represents the state register storing the initial condition of the differential equation $\ket{\psi_0}$.
    \item \textbf{Algorithm:} The time interval $[0, T]$ is then divided into a series of short segments (time steps), separated by temporal mesh points $0 = t_0 < t_1 < \cdots < t_L = T$ with $t_k = \tau k$ and $\tau = T/L$. For each time step, the following two steps are performed:
    \begin{itemize}
        \item \textbf{Step 1} (Unitary Evolution): Apply the unitary dynamics governed by the Hamiltonian $ H$, specifically $ e^{-iH \tau}$, to the system register. 
        \item \textbf{Step 2} (Purification and Postselection): 
        Apply $ e^{ i \sqrt{2\tau}  G_j}$ sequentially to both the purification and system registers for $j$ from $1$ to $J$. After each application, measure the purification register; if the outcome is $0$, proceed to the next $ e^{ i\sqrt{2\tau}  G_j}$. If any measurement outcome is non-zero, discard the realization.
    \end{itemize}
\item \textbf{Output:} For the successful realization, we get $\ket{0}\ket{\widetilde{\psi}(T)}$ (up to normalization factor), where $\ket{\widetilde{\psi}(T)}$ is an unnormalized vector representing the approximate solution of the differential equation $\ket{\psi}$ as desired. The procedure can be repeated to prepare multiple states to obtain observable expectations.
\end{itemize}

\begin{figure*}[t]
\centering
    \adjustbox{max width=2\linewidth}{
        \centering
    \begin{quantikz}
         \lstick{$\ket{0}$}& \qw & \qw &
        \gate[2]{e^{i\sqrt{2\tau}G_1}}\gategroup[2,steps=3,style={inner
sep=6pt}]{} & \meter{} & 
        \qw  & \ldots  \ldots  & \gate[2]{e^{i\sqrt{2\tau}G_J}}\gategroup[2,steps=3,style={inner
sep=6pt}]{} & \meter{} & \qw & \qw\\
         \lstick{$\ket{\psi_0}$} & \qwbundle{n} & \gate{e^{-iH\tau}} & \qw & \qw & \qw & \ldots  \ldots & \qw & \qw & \qw & \qw
    \end{quantikz}
    }
    \caption{Quantum Circuit Diagram of the ODE algorithm described above implemented for a single timestep $\tau$. Note that the circuit is for an operator $A \equiv - iH  -\sum_{i=1}^J L_i^{\dagger}L_i$ containing $J$ Hermitian terms and that all nonzero measurement results for the ancilla qubit are discarded. $\ket{\psi_0}$ is also over all $n$ qubits of the Hamiltonian $H$.}
    \label{fig:circuit_diagram}
\end{figure*}

The operator we implement in one time step is
\begin{equation}
    \label{eq:single_time_step_ODE}
    M(\tau) = (I\otimes e^{-i\tau H})(\ket{0}\bra{0}\otimes I) e^{-i\sqrt{\tau} G_J} 
    \cdots (\ket{0}\bra{0}\otimes I) e^{-i\sqrt{\tau} G_2}(\ket{0}\bra{0}\otimes I) e^{-i\sqrt{\tau} G_1}.
\end{equation}
From the above discussion, we can see that
\begin{equation}
\label{eq:high_precision_limit_solution}
    \ket{\psi(T)}=e^{-A T}\ket{\psi_0} = M(T/R)^R \ket{\psi_0} + \Or(T^2/R),
\end{equation}
where $R>0$ is an integer representing the number of time steps, each of length $\tau=T/R$. Therefore, we obtain the solution of the ODE by repeatedly applying the operator $M(T/R)$, which consists of unitary operations, measurements, and post-selection.

We note that although the algorithm involves post-selection, it does not pay a success probability loss due to the subnormalization (operator norm of the operators) as typically required in the block-encoding-based procedures. Instead, it depends only on the state.
This can be made transparent by estimating the success probability of this procedure.

\REV{Although here we focus on the case of using only one ancilla qubit for the entire algorithm, it is very easy to modify the algorithm to use multiple ancilla qubits, implementing multiple dissipative terms in parallel, which can help reduce the circuit depth. Using one ancilla qubit also faces the issues of unfavorable error propagation and idling errors accumulating on qubits not being used. In the end the user should choose the optimal number of ancilla qubits to use based on the number of available qubits and circuit depth.}

\subsection{The Success Probability}
\label{sec:success_prob}

Because of the non-trace-preserving nature of the dynamics when seen from a density matrix perspective, we do not expect to have an efficient algorithm with large success probability for arbitrary $A$ and for large time $T$. Such an algorithm would enable efficiently post-selecting measurement results, thus violating complexity-theoretic lower bounds (e.g., \cite[Theorem~10]{FangLinTong2023}). Rather, in this work we aim for the \emph{natural success probability} of the problem that results from the decay of the trace of the density matrix.

The use of post-selection raises the question of a diminishing success probability as the algorithm proceeds.  However, we can compute this success probability to be $\|M(t/R)^R\ket{\psi_0}\|^2$, which converges to $\|\ket{\psi(t)}\|^2$ as $R\to \infty$. Therefore we recover the natural success probability of this problem.

\subsection{Preserving Locality}
\label{sec:locality}
We note that our algorithm has the nice feature that it is \emph{locality-preserving}. Let us consider the case where
\begin{equation}
\label{eq:Pauli_decomp_ham_jump}
    H = \sum_a h_a P_a,\quad L_j = \sum_{\REV{b}} \lambda_{j,b}P_b,
\end{equation}
where each $P_a$ is a Pauli operator. In physical applications these Pauli operators are typically low-weight, i.e., they only act non-trivially on a few qubits. A concrete example is given in~\cref{sec:appl_Hatano-Nelson}. 

In our algorithm, we only need to implement $e^{-iH\tau}$ and $e^{-i\sqrt{\tau}G_j}$ for each $j$. For $e^{-iH\tau}$, further Trotterization reduces it to implementing $e^{-iP_a\tau}$ for each $P_a$, which is the most basic operation needed in almost all Hamiltonian simulation algorithms. For $G_j$, we can in fact write it as a linear combination of Pauli operators through \eqref{eq:defn_Gj} and \eqref{eq:Pauli_decomp_ham_jump}:
\begin{equation}
    G_j = \sum_b \Re\lambda_{j,b}X\otimes P_b + \sum_b \Im\lambda_{j,b}Y\otimes P_b,
\end{equation}
where the extra Pauli-$X$ and $Y$ operators act on an ancilla qubit.
From the above we can see that $e^{-i\sqrt{\tau}G_j}$ can likewise be decomposed into time evolutions generated by Pauli operators, whose weight is greater than the original weight by one. Therefore when the problem involves $H$ and $L_j$ that are $k$-local, then our algorithm only need to implement unitary time evolutions generated by at most $(k+1)$-local Pauli operators. In this sense we say our algorithm is locality preserving. This is in contrast with algorithms based on linear combination of unitaries, where the extensive control structure can greatly increase the weight of the Pauli operators that need to be implemented.

\section{Connection to Lindbladian Evolution}
\label{sec:connection_to_lindbladian}

We will connect solving the ODE \eqref{eq:ode_general} to solving a Lindblad master equation 
\begin{equation}
    \frac{\rd}{\rd t}\rho(t) = A\rho+\rho A^\dag + \sum_j L_j\rho L_j^\dag.
\end{equation}
One can see from the above that if we can get rid of the term $\sum_j L_j\rho L_j^\dag$ on the right-hand side then we will recover the equation governing the time evolution of $\ket{\psi(t)}\bra{\psi(t)}$, which, if we write $\rho(t)=\ket{\psi(t)}\bra{\psi(t)}$, is
\[
 \frac{\rd}{\rd t}\rho(t) = A\rho+\rho A^\dag .
\]
From this perspective, the algorithm we discussed in~\cref{sec:algorithm} can be derived from post-selecting an algorithm for Lindbladian simulation.

A common way to simulate the Lindbladian dynamics is to introduce an ancilla qubit and dilate some of the terms~\cite{ChildsLi2016, CleveWang2017, WocjanTemme2023, ChenKastoryanoBrandaoGilyen2023, ChenKastoryanoGilyen2023, LiWang2023, DingLiLin2023,PocrnicSegalWiebe2024,LiWang2022,HeLiLiLiWangWang2024}. More precisely, we use the $G_j$ operators defined in \eqref{eq:defn_Gj}.
When simulating the Lindbladian dynamics,  at each time step, we will reset the ancilla qubit to $\ket{0}$, implement the unitary $e^{-i\sqrt{\tau}G_j}$ where $\tau$ is the time step size, and when we trace out the ancilla qubit, which does not require any physical operation, we will be implementing the Lindbladian terms corresponding to the jump operator $L_j$. Doing the same for each jump operator, and implementing the unitary time evolution $e^{-i\tau H}$, we will then have a single time step for the Lindbladian dynamics, which can be repeated to simulate long-time evolution.

To obtain the ODE solution, only a small modification is needed: instead of tracing out the ancilla qubit, we measure it and post-select the measurement result. We only proceed when the measurement returns the $\ket{0}$ state on the ancilla qubit. 
The action on the quantum state can then be described by the operator on the left-hand side of \eqref{eq:first_order_approx_dissipative_evolution}, which approximates the evolution $e^{-\tau L_j^\dag L_j}$ up to first order.
This then enables us to simulate a time step of the ODE. 

\REV{Effectively, our algorithm realizes a continuously monitored quantum dynamics, which has been studied in quantum optics \cite{Carmichael2008statistical,Brun2002simple}. The idea of using an ancilla qubit to implement dissipative dynamics has also been considered in many settings, such as to stabilize a bosonic qubit \cite{RoyerSinghEtAl2020stabilization}.}

\section{Error Analysis and Runtime}
In this part, we omit the subscript $j$ for notational simplicity. According to the Taylor theorem, we have 
\begin{equation}
    e^{i\sqrt{2\tau} G} = I + i\sqrt{2\tau} G - \tau G^2 - \frac{i\sqrt{(2\tau)^3}}{3!} G^3 
        + \int_0^{\sqrt{2\tau}}\frac{G^4}{3!} e^{i \alpha G} (\sqrt{2\tau} - \alpha)^3 \rd\alpha,
\end{equation}
and a direct computation reveals that 
\begin{equation}
      \left( \bra{0} \otimes I \right) 
    e^{ i \sqrt{2\tau}  G} \ket{0} \ket{\psi}
      = (I - \tau L^\dagger L)\ket{\psi} 
            + \left( \bra{0} \otimes I \right)  
    \int_0^{\sqrt{2\tau}}\frac{G^4}{3!} e^{i \alpha G} (\sqrt{2\tau} - \alpha)^3 \rd\alpha \ket{0} \ket{\psi}.
\end{equation}
Similarly by the Taylor theorem, we have
\begin{equation}
    e^{- \tau L^\dagger L} = I - \tau L^\dagger L + \int_0^\tau (L^\dagger L)^2 e^{-\alpha L^\dagger L} (\tau-\alpha) \rd \alpha.
\end{equation}
Therefore, the error can be estimated via the norm of the remainder terms in the Taylor expansion
\begin{equation}
\begin{aligned}
    &\norm{\left( \bra{0} \otimes I \right) 
    e^{ \sqrt{2\tau}  G} \ket{0} \ket{\psi}- e^{\tau L^\dagger L} \ket{\psi}} 
    \\
    \leq & \norm{(L^\dagger L)^2 \ket{\psi}} \left(\frac{\tau^2}{6} + \frac{\tau^2}{2}\right) = \norm{(L^\dagger L)^2 \ket{\psi}} \frac{2\tau^2}{3} ,
\end{aligned}
\end{equation}
where we used the fa\textbf{}cts that 
\begin{equation}
    \norm{e^{i\alpha G}} = 1, \quad \norm{e^{-\alpha L^\dagger L}} \leq 1.
\end{equation} 
Besides this error, the algorithm also has an error coming from the Trotterization
\begin{equation}
\norm{e^{ A_J \tau} \cdots e^{ A_1 \tau} e^{ A_0 \tau} -  e^{ A \tau}} \leq \frac{1}{2} \sum_{j=0}^J\norm{[ \sum_{k = 0}^{j-1} A_k, A_j]} \tau^2,
\end{equation}
where $A_0 = -iH$ and $A_j = -L^\dagger_j L_j$. Combining both errors, in $R$ time steps, each with length $\tau=T/R$, we have the cumulative error bound
\begin{equation}
\begin{aligned}
    & \norm{\ket{\psi(T)} - \ket{\tilde\psi(T)}} 
    \\
    \leq &  \frac{1}{2} \sum_{j=0}^J\norm{[ \sum_{k = 0}^{j-1} A_k, A_j]} \sup_{t \in [0,T]} \norm{\ket{\psi (t)}} \frac{T^2}{R} 
 + \sum_j\sup_{t \in [0,T]}\norm{(L_j^\dagger L_j)^2 \ket{\psi (t)}} \frac{2 T^2}{3R}.
\end{aligned}
\end{equation}

\begin{lemma}
\label{lem:normalized_err}
If $\|\ket{\psi}-\ket{\phi}\|\leq \frac{1}{2}\|\ket{\psi}\|$, then 
\[
\left\|\frac{\ket{\psi}}{\|\ket{\psi}\|}-\frac{\ket{\phi}}{\|\ket{\phi}\|}\right\| \leq \frac{4\|\ket{\psi}-\ket{\phi}\|}{\|\ket{\psi}\|}.
\]
\end{lemma}
Thanks to \cref{lem:normalized_err}, to achieve $\epsilon$ for the normalized vectors, namely,
\begin{equation}
    \left\|\frac{\ket{\psi}}{\|\ket{\psi}\|}-\frac{\ket{\phi}}{\|\ket{\phi}\|}\right\| \leq \epsilon,
\end{equation} it is sufficient to choose the number of time steps $R$ such that 
 \begin{equation}
      \norm{\ket{\psi(T)} - \ket{\tilde\psi(T)}} = \Theta \left(\epsilon \norm{\ket{\psi(T)}} \right), 
 \end{equation}
 which yields 
 \begin{equation}
 \label{eq:num_time_steps}
     R = \max \Bigg\{ \frac{\sum_{j=0}^J\norm{[ \sum_{k = 0}^{j-1} A_k, A_j]} \sup_{t \in [0,T]} \norm{\ket{\psi (t)}} T^2}{ \norm{\ket{\psi(T)}} \epsilon},
          \frac{ \sum_j \sup_{t \in [0,T]}\norm{(L_j^\dagger L_j)^2 \ket{\psi (t)}}  T^2}{\norm{\ket{\psi(T)}} \epsilon}
     \Bigg\}.
 \end{equation}
Without loss of generality, we let $\norm{\ket{\psi_0}} = 1$. Upon measuring the purification qubits and getting measurement outcomes all 0, we get the state $\ket{\tilde\psi}$ that is an approximate solution. The success probability of it is at least 
 \begin{equation}
 \norm{ \left( \prod_{j = 1}^J \left( I - \tau L_j^\dagger L_j \right) e^{-iH \tau} \right)^R\ket{\psi_0} }^2  = \norm{\ket{\tilde \psi(T)}}^2.
 \end{equation}
 Note that $\norm{\ket{\tilde \psi (T)}} = (1 + \Theta(\epsilon)) \norm{\ket{\psi (T)}} $, and hence the success probability is $\Omega(\norm{\ket{\psi(T)}}^2 )$. Generally, if we do not let $\norm{\ket{\psi_0}} = 1$, the total success probability is 
 \begin{equation}\label{eq:def_q}
     \frac{1}{q^2}: = \frac{\norm{\ket{\psi(T)}}^2}{\norm{\ket{\psi_0}}^2},
 \end{equation}
 where the parameter $q$ denotes the state ratio that appears in all quantum ODE solvers.
 Therefore, the number of times that $e^{i\sqrt{\tau}G}$ and $e^{-iH\tau}$ need to be applied (for $t>0$) is
\begin{equation}
\label{eq:query_complexity}
     q^2 R = \frac{\norm{\ket{\psi_0}}^2}{\norm{\ket{\psi(T)}}^3} \max 
     \Bigg\{ 
     \sum_j \sup_{t \in [0,T]}\norm{(L_j^\dagger L_j)^2 \ket{\psi (t)}} ,
               \sum_{j=0}^J\norm{[ \sum_{k = 0}^{j-1} A_k, A_j]}\sup_{t \in [0,T]} \norm{\ket{\psi (t)}} 
     \Bigg\} \frac{  T^2}{  \epsilon}.
\end{equation}
\begin{theorem}
    \label{thm:query_complexity}
    A state $\ket{\tilde{\psi}(T)}$ that satisfies $\|\ket{\tilde{\psi}(T)}-\ket{\psi(T)}\|\leq \epsilon\|\psi(T)\|$, can be prepared up to a normalization factor using a number of applications of $e^{-iH\tau}$ and $e^{-i\sqrt{\tau}G_j}$ given in \eqref{eq:query_complexity}.     The number of calls needed for the initial state preparation is $\Or(q^2)$. Here $\psi(t)$ denotes the solution to the ODE \eqref{eq:ode_general}, and $\tau=T/R$, with $R$ given in \eqref{eq:num_time_steps}.
\end{theorem}

We note that the algorithm, when implemented directly without advanced post-processing techniques, achieves first-order accuracy.

Here, the number of applications needed for the initial state preparation is $\mathcal{O}(q^2)$, which does not depend on the precision $\epsilon$. This marks a clear improvement over prior QLSA-based differential equation solvers, where the state preparation cost also depends on $\epsilon$. While our $\mathcal{O}(q^2)$ cost is not optimal compared to the optimal $\mathcal{O}(q)$ scaling achieved in~\cite{FangLinTong2023,AnLiuLin2023linear,AnChildsLinYing2024laplace,AnChildsLin2023quantum,ShangGuoAnZhao2024,LowSu2024} using more advanced fault-tolerant subroutines, it is already favorable due to its independence from precision.

Achieving the optimal $\mathcal{O}(q)$ state preparation would require applying amplitude amplification. Although our near-term algorithm includes mid-circuit measurements, these can be deferred to yield a coherent implementation suitable for amplitude amplification. However, doing so would necessitate more ancilla qubits, which is not desirable in our setting. Therefore, we opt for the $\mathcal{O}(q^2)$ version here, which is qubit-efficient while still offering competitive performance. Notably, for algorithms that also avoid amplitude amplification, our state preparation cost asymptotically matches that of other state-of-the-art fault-tolerant algorithms that require a large number of qubits.

We will next analyze the gate complexity of the algorithm when applied to a geometrically local $A$. More precisely, we assume that $H$ and each $L_j$ are geometrically local, i.e., their Pauli decompositions described in \eqref{eq:Pauli_decomp_ham_jump}, satisfy that each Pauli operator $P_a$ or $P_b$ is supported on $\mathsf{k}=\Or(1)$ adjacent qubits arranged on a constant dimensional lattice of $n$ qubits, $J=\Or(n)$, and all coefficients are contained in $[-1,1]$. With these assumptions, we can see that each $A_j$ commutes with all but $\Or(1)$ other $A_k$, for $1\leq j,k\leq J$. $A_0=-iH$ may fail to commute with all the $A_j$, but $\|[A_0,A_j]\|=\Or(1)$. Based on this observation, we have
\[
\sum_{j=0}^J\norm{\left[ \sum_{k = 0}^{j-1} A_k, A_j\right]} = \Or(n).
\]
For $L_j^\dag L_j\ket{\psi(t)}$ we also have
\[
\|L_j^\dag L_j\ket{\psi(t)}\|\leq \|L_j\|^2 \|\psi_0\|=\Or(\|\psi_0\|),
\]
where we have used the decay of the solution norm in \eqref{eq:norm_decay}. Therefore by~\cref{thm:query_complexity}, the number of applications of $e^{-iH\tau}$ and $e^{-i\sqrt{\tau}G_j}$ is $\Or(q^3 n T^2/\epsilon)$.

Since our final goal is the gate complexity, we want to estimate the number of gates needed to implement $e^{-iH\tau}$ and $e^{-i\sqrt{\tau}G_j}$. Since $G_j$ only involves a constant number of qubits, $e^{-i\sqrt{\tau}G_j}$ can be implemented with $\Or(1)$ single- and two-qubit gates. For $e^{-iH\tau}$, implementing it requires further Trotterization and will introduce further error. However, due to the geometrically local assumption, the step size selection through \eqref{eq:num_time_steps} is sufficient to control the error introduced by Trotterizing $e^{-iH\tau}$ as well. In each time step decomposing $e^{-iH\tau}$ using the first-order Trotter formula introduces an error of order $\Or(n\tau^2)$ according to \cite[Eq.~(119)]{ChildsSuTranEtAl2020}. Note that this error is then on the same order as that introduced by Trotterizing $A_0,A_1,\cdots,A_J$, and therefore we can control the relative error to be at most  $\epsilon$ by simply halving the step size. From the above discussion, we can see that we only need to multiply $\Or(n)$ to the number of applications of $e^{-iH\tau}$ and $e^{-i\sqrt{\tau}G_j}$ to obtain the gate complexity, which is $\Or(q^3 n^2 T^2/\epsilon)$.

\begin{corollary}
    \label{cor:geometrically_local}
    For geometrically local $H$ and $L_j$, the ODE \eqref{eq:ode_general} with coefficient matrix $A=-iH-\sum_j L_j^\dag L_j$ on $n$ qubits can be solved with gate complexity $\Or(q^3 n^2 T^2/\epsilon)$, producing a state $\ket{\tilde{\psi}(T)}$ up to a normalization factor that satisfies $\|\ket{\tilde{\psi}(T)}-\ket{\psi(T)}\|\leq \epsilon\|\psi(T)\|$. In the above $q=\|\psi_0\|/\|\psi(T)\|$.
\end{corollary}

\REV{According to the analysis in this section, our method requires a quantum circuit of depth $\Or(q n^2 T^2/\epsilon)$ repeated $\Or(q^2)$ times. With $\Or(n)$ ancilla qubits instead of one, the circuit depth can be reduced to $\Or(q n T^2/\epsilon)$ by implementing multiple dissipative terms in parallel. In terms of error analysis, a parallel implementation of multiple dissipative terms is equivalent to a certain ordering in a sequential implementation. Since our analysis does not depend on the ordering of the terms, it will remain unchanged with this parallel implementation.} 

\REV{A $q$ dependence appears in the circuit depth because we need to make the time step size $\tau=\Or(\epsilon/(Tq))$ so that the cumulative relative error in the end is below $\epsilon$. We remark that this may very well be an over-estimation, since in the analysis we did not take into account the decay of the error. More precisely, the dissipative nature of the dynamics can cause the error to decay as well, and an ``average case'' error accumulation in multiple steps should be sublinear as opposed to the worst-case linear accumulation used in the above analysis. Empirically, as demonstrated in the numerical example in~\cref{sec:numerics}, step sizes much larger than $1/q$, thus making the circuit depth independent of $q$, can still yield qualitatively correct results.}

\REV{One can directly apply amplitude amplification~\cite{BrassardHoyerMoscaEtAl2002} to improve the $q$ dependence in our algorithm, but this comes at the cost of requiring additional ancilla qubits and much larger circuit depth. In the numerical example studied in~\cref{sec:numerics}, with a success probability around $10^{-3}$, amplitude amplification would require a circuit depth that is more than $10$ times as large, making the simulation much more sensitive to noise. Therefore, in this work, we focus on the algorithm without amplitude amplification, while improvement in the size and fidelity of available quantum hardware can make the amplitude amplification procedure beneficial in the future.}

\section{Application to the Interacting Hatano-Nelson Model}\label{sec:appl_Hatano-Nelson}
\par The ability to solve for the time evolution of a general matrix $A$ allows for the consideration of the time evolution via the Schr\"odinger equation of a much broader class of Hamiltonians that are not Hermitian. These non-Hermitian Hamiltonians have recently become an increasing area of interest in quantum science due to their ability to capture novel kinds of behavior not observable with Hermitian Hamiltonians. 
Mid-circuit measurements along with post selection have previously been used in the simulation of small non-Hermitian Hamiltonians \cite{ZhangEtAlNature2021,JebraeilliEtAlPRA2025}. However, these methods provided no rigorous performance guarantees. Our ODE method is generalizable for large non-Hermitian Hamiltonians so long as the non-Hermitian component can be converted into a PSD form. One such non-Hermitian Hamiltonian that allows for a scaled implementation with our algorithm is the interacting Hatano-Nelson model. 

\subsection{The Model and the Algorithm}
Non-Hermitian Hamiltonians have many unique features, one of which is the admission of exceptional points. These are critical points where the Hamiltonian is no longer diagonalizable~\cite{Heiss2012}. Another particularly important unique behavior of non-Hermitian Hamiltonians, and one which we focus on for our numerical experiments, is the non-Hermitian skin effect. This effect characterizes a localization of an extensive number of eigenstates of the non-Hermitian Hamiltonian and arises from the non-reciprocity due to the non-Hermitian component of the Hamiltonian~\cite{KawabataNumasawaPRX2023, LeePRL2016}. The skin effect plays an important role in the topological phases that arise for the non-Hermitian Hamiltonian \cite{RudnerLevitov2009}. 
\par A prototypical example of a non-Hermitian Hamiltonian that admits all of the above behaviors is the Hatano-Nelson (HN) model \cite{NelsonHatanoPRL1996, NelsonHatanoPRL1997, NelsonHatanoPRL1998}.
The model can be defined for a 1D lattice of fermions or bosons with the open boundary condition (OBC), and we will focus on the fermionic version. The spinless Hamiltonian for this many-body HN model with OBC is given below: 
\begin{align}
     H_{HN} = \sum_{j=1}^{N-1} \left(  (J+ \gamma)c^{\dagger}_{j+1}c_{j} + (J- \gamma)c^{\dagger}_{j}c_{j+1}\right)
 \end{align}
 Where $c_j, c_j^{\dagger}$ are spinless fermionic annihilation and creation operators and $J,\gamma \in \mathbb{R}$ represent the symmetric Hermitian and asymmetric non-Hermitian terms, respectively.

While the HN model is quadratic in the fermionic creation and annihilation operators and is therefore easy to classically simulate, we will consider a modified version that involves many-body interactions. Such a model allows us to study the stability of the non-Hermitian skin effect and entanglement phase transition under many-body perturbation, and has a richer phenomenology such as $\mathcal{PT}$-transition \cite{ZhangDennerEtAl2022symmetry}.
The Hamiltonian for the nearest-neighbor interacting Hatano-Nelson model is given below: 
\begin{equation}
\label{eq:interacting_hatano_nelson}
    \begin{aligned}
    H_{HN}  &= \sum_{j=1}^{N-1} H_{HN, j} + V
    \\
    &=  \sum_{j=1}^{N-1}J(c^{\dagger}_{j+1}c_{j} + c^{\dagger}_{j}c_{j+1})
 + \gamma(c^{\dagger}_{j+1}c_{j}-c^{\dagger}_{j}c_{j+1})  + \sum_{j=1}^{N-1}V_{j,j+1} n_j n_{j+1}, 
\end{aligned}
\end{equation}
 where the first and second terms of $H_{HN,j}$ represent the Hermitian and anti-Hermitian components of the non-interacting component of the Hamiltonian, respectively, and $V$ is the added quartic interaction. For simulation on quantum computers, we convert each $H_{NH, j}$ from a fermionic representation to one composed of spin-1/2 Pauli operators. We use the Jordan-Wigner transform to do so, yielding the following representations for $H_{HN, j}$: 
\begin{equation}
\begin{aligned}
     H_{HN,j} = &\frac{J}{2} (Y_jY_{j+1} + X_jX_{j+1}) - \frac{i\gamma}{2}(Y_jX_{j+1} -X_{j}Y_{j+1})
     \\ & = H_{H,j} + H_{A,j}, 
\end{aligned}
\end{equation}
where we have again separated each Hamiltonian into Hermitian ($H_{H,j}$) and anti-Hermitian ($H_{A,j}$) terms. 
For the interaction term $V$, it is transformed to
\begin{equation}
    V = \frac{1}{4}\sum_{j}V_{j,j+1}(I-Z_j)(I-Z_{j+1}).
\end{equation}
Looking at the Schr\"odinger equation from the perspective of the ODE in \eqref{eq:ode_general}, $H_{NH}$ is related to $A$ in the following way:
\begin{align}
   -iH_{NH} - (N-1)c I = A = -iH - \sum_jL^{\dagger}_j L_j,
\end{align}
where
\begin{equation}
    H = \sum_{j=1}^{N-1} H_{H,j} + V,\quad L_j^\dag L_j = iH_{A,j} +c I.
\end{equation}
Here a shift $cI$ is needed to ensure the positive semidefiniteness of $L_j^\dag L_j$.
The component $-iH$ is trivially simulated by 
Trotterization. We therefore only focus on the anti-Hermitian component when realizing the algorithm. Before we can apply the algorithm, we need to determine the shift $c$. To do this we find the minimum eigenvalue of $iH_{A,j}$ and subtract it from $iH_{A,j}$. Thus yielding the new operator $K_j$:
\begin{align}
    K_j = iH_{A,j} + \gamma I = \frac{\gamma}{2}(Y_jX_{j+1} -X_jY_{j+1}) +  \gamma I.
\end{align}
In other words we set $c=\gamma$.
To find $L_j$ such that $K_j=L_j^\dag L_j$, we can simply choose $L_j=\sqrt{K_j}$, which is well-defined because $K_j$ is positive semidefinite.
Solving for this yields: 
\begin{equation}
    L_j = \frac{\sqrt{\gamma }}{2}\Bigg( \left( 1 - \frac{1}{\sqrt{2}}\right)Z_jZ_{j+1} 
        + \frac{1}{\sqrt{2}}(Y_jX_{j+1} - X_jY_{j+1})  + \left( 1 + \frac{1}{\sqrt{2}}\right)I \Bigg)
\end{equation}
Given the Pauli string description of $L_j$, we can construct $G_j$ given in \eqref{eq:defn_Gj}. Because $L = L^{\dagger}$, we have
\begin{equation}
    \begin{aligned}
    &G_j = X_0 L_j \\
    = & \frac{\sqrt{ \gamma }}{2}\Bigg( \left( 1 - \frac{1}{\sqrt{2}}\right)X_0Z_jZ_{j+1} + \left( 1 + \frac{1}{\sqrt{2}}\right)X_0I_{j}I_{j+1}
        + \frac{1}{\sqrt{2}}(X_0Y_jX_{j+1} - X_0X_jY_{j+1})  \Bigg)
\end{aligned}
\end{equation}
Note that the ancilla qubit is positioned at index 0. 

From the above discussion, we can see that our algorithm can simulate the real-time dynamics of the interacting Hatano-Nelson model in \eqref{eq:interacting_hatano_nelson}. The locality of the $H$ and $G_j$ coming from the Jordan-Wigner transform ensures that the simulation can be done with gate complexity given in~\cref{cor:geometrically_local}.

\REV{We can compare the performance of our algorithm with previous proposals for quantum ODE solvers such as \cite{FangLinTong2023,ChildsLiu2019,Krovi2022,BerryCosta2024}. All of these methods have better asymptotic scaling in terms of the gate complexity compared to our algorithm. However, to the best of our knowledge, only small-scale demonstrations with dimensions at most $4$ on classical simulators have been made for some of these algorithms~\cite{Classiq-time-marching,Classiq-Lanchester}. This is partly due to the high non-asymptotic gate complexity needed to implement complicated quantum subroutines even for small problems. By comparison, we include numerical simulation results for our algorithm for the interacting Hatano-Nelson model in~\cref{sec:numerics}.}

\REV{It is much easier to compare the number of qubits needed for the simulation. For concreteness, we consider the setting of a $7$-site interacting Hatano-Nelson model simulated for 10 steps. There are $7$ Pauli operators for each pair of adjacent sites, making the total number of Pauli operators in the Hamiltonian $42$. Therefore at least $\lceil\log_2(42)\rceil=6$ ancilla qubits are needed for block-encoding the Hamiltonian alone. The algorithm in \cite{FangLinTong2023} requires keeping track of the success of each step coherently, which requires $\lceil\log_2(10)\rceil=4$ ancilla qubits for this purpose. The algorithms in \cite{BerryCosta2024,ChildsLiu2019,Krovi2022} are all based on solving a linear system involving all time steps ($10$ in our example), which requires $\lceil\log_2(10)\rceil=4$ additional ancilla qubits to store the linear system. From the above we can see that these algorithms can easily require more than $10$ ancilla qubits for this simulation, and this is an underestimation because the additional ancilla qubits needed for arithmetic operations, matrix exponentials, and amplitude amplification are not counted. In contrast, our algorithm requires $1$ ancilla qubits even though one can choose to use more for better parallelizability.}

\subsection{Numerical Experiments}
\label{sec:numerics}

\REV{To demonstrate the near term utility of this algorithm for solving for the time dynamics of non-Hermitian Hamiltonians, we implement it as a quantum circuit using the Qiskit SDK \cite{JavadiEtAl2024quantum} according to~\cref{fig:circuit_diagram}. It is desired to use the evolution circuit for the interacting Hatano-Nelson model described in the previous section to observe a phenomenon unique to non-Hermitian Hamiltonians. The non-Hermitian skin effect is a prototypical example of this. A salient feature of the non-Hermitian skin effect is the accumulation of particles towards one boundary with an open boundary condition. This effect can be observed by measuring the occupation number at each site.}

\begin{figure}[h!]
    \centering
    \includegraphics[width=0.48\linewidth]{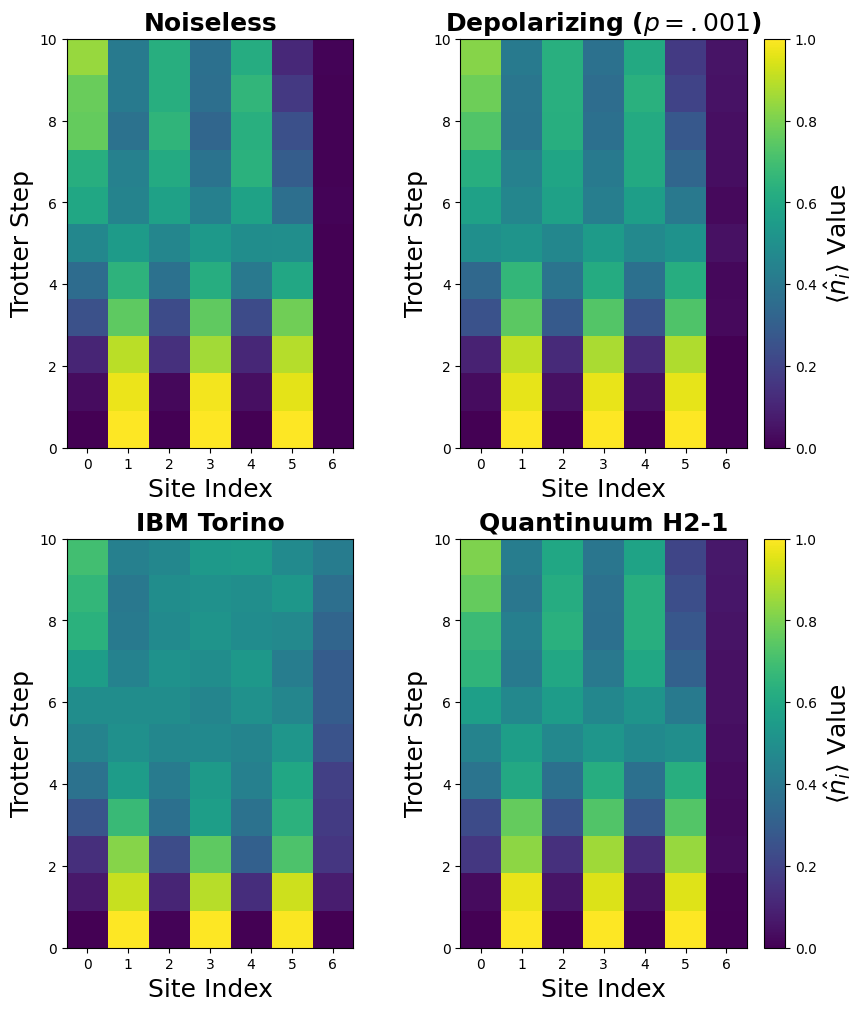}
    \caption{\REV{Site fermionic occupation time dynamics for the interacting Hatano-Nelson model in \eqref{eq:interacting_hatano_nelson} $(J = 1.0, \gamma = 0.8, V_{i,i+1} = 2.0, \tau = 0.1)$. The evolution was computed using a quantum circuit-level simulation of the algorithm. The quantum circuit for the multi-step time evolution was simulated noiselessly, for single qubit depolarizing noise $p=.001$, a simulated IBM Torino Backend, and an approximate noise model emulator for the Quantinuum H2-1 processor. The initial state is given by the fermionic state $\ket{\psi_0} = \ket{0101010}$, with all odd sites occupied. The quantum circuit for a single time step implemented required 214 single and two-qubit gates once compiled}}
    \label{fig:skineffect}
\end{figure}

\REV{To do this, we implement the time evolution for the interacting Hatano-Nelson model for multiple Trotter steps ranging from 0-10 and then measure all the qubits corresponding to the lattice sites of the interacting Hatano-Nelson model. After applying the post-selection procedure required by the algorithm, the fermionic occupation number $\braket{\hat{n}_i}$ of each site is found by measuring the probability of measuring 0 at each lattice site.  The measured occupation number is then plotted on a 2D space-time color plot as shown in~\cref{fig:skineffect}.  The resulting simulations shown in~\cref{fig:skineffect} clearly show the non-Hermitian skin effect, with the fermionic occupation localization on the left boundary of the lattice for all of the simulations.}

\REV{We note that an implementation of the interacting Hatano-Nelson model has been done on superconducting qubits in \cite{ShenChenEtAl2025observation}, where the authors applied circuit optimization and error mitigation techniques to clearly observe the non-Hermitian skin effect. Their implementation of the non-Hermitian Hamiltonian term is based on encoding the non-unitary operator into a larger unitary operator, which is constructed through QR decomposition and circuit synthesis (see the discussion around \cite[Eq.~(21)]{ShenChenEtAl2025observation}). This can lead to unfavorable gate count in practice and results in an exponential runtime when the non-unitary operator is non-local (e.g., for the convection-diffusion equation in \eqref{eq:convection_diffusion}). By comparison we only need to implement Pauli rotations that are directly computed from the non-Hermitian Hamiltonian for the interacting Hatano-Nelson model. Moreover, in this work we are proposing a general algorithm for linear ODEs with the interacting Hatano-Nelson model only as an application.}

\begin{figure}[h!]
    \centering
    \includegraphics[width=0.6\linewidth]{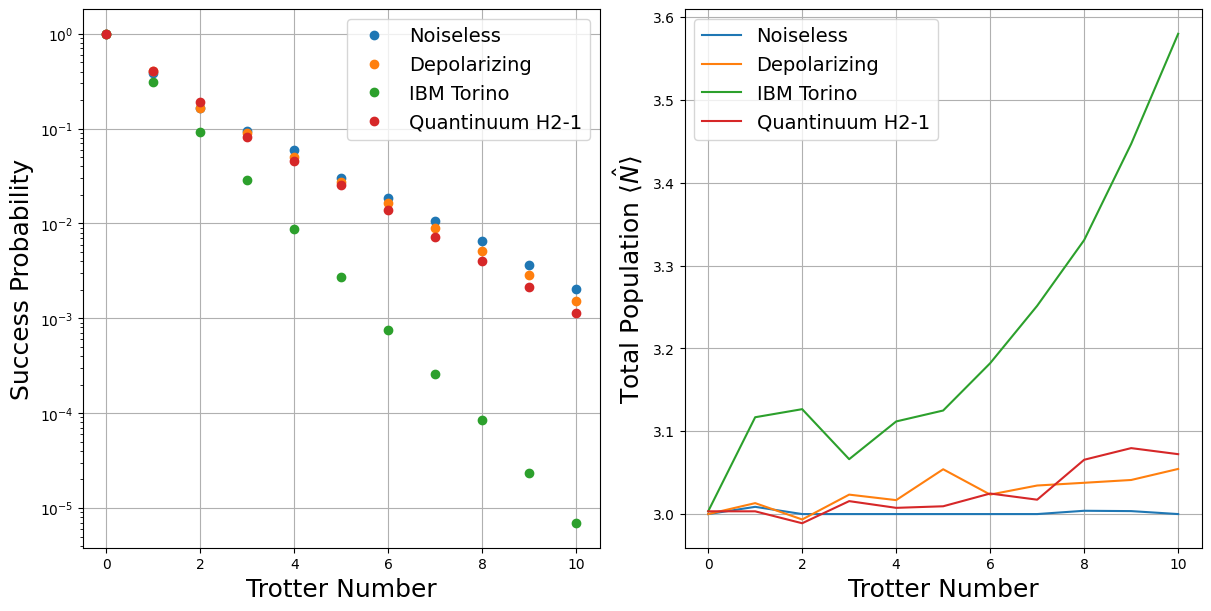}
    \caption{\REV{(Left) Success probability for the algorithm implementing a multi-time step evolution for the interacting HN model $(J = 1.0, \gamma = 0.8, V_{i,i+1} = 2.0, \tau = 0.1)$. Success probability was computed by measuring the fraction of total number of runs at the end of the time steps for which the ancilla qubit measurements were measured to be all 0. (Right) Plotted total fermionic occupation number $\hat{N}$ for the lattice over time.}}
    \label{fig:sucprob}
\end{figure}

\REV{
The success probability at each Trotter step was computed via the number of successful post-selected shots and is given in \cref{fig:sucprob}. The total population $\hat{N}$ of the lattice was also measured and served as an important metric for the reliability of the simulations as it is a conserved quantity for the interacting Hatano-Nelson model and therefore its measured value should remain constant. To improve the simulation time of computing the time evolution and ensure enough valid post-selected shots, the shot count was scaled exponentially in the number of time steps to compensate for the approximate exponential decay in the measured success probability. The initial shot count was set at 300 and scaled by a reciprocal decay probability constant raised to the Trotter number ($N = \frac{300}{\rho^T}$, where $\rho$ is the estimated success probability at each step). This was truncated at 10,000,000 shots to ensure the simulation could be performed in a reasonable time frame on a laptop ( $<$10Hrs).}

\REV{
To evaluate the performance of the quantum circuit on hardware, 4 different circuit level simulations were performed: A noiseless state vector simulation, a noisy one with depolarizing error $p =.001$ applied to each qubit, a noisy simulation emulating the IBM Torino quantum processor, and a custom noisy simulation that emulates the behavior of the Quantinuum H2 processor. The Quantinuum noise model was constructed in Qiskit using publicly available noise parameters that included T1 and T2 times, measurement errors, and 1 and 2 qubit gate errors. All of these simulations were performed on a Dell XPS 15 with an Intel i9 CPU.}

\REV{Some special treatment was needed to implement the dynamical circuit so that the simulation can be terminated upon an unsuccessful measurement.
Because of the post-selection requirements for the algorithm along with the classical control flow constraints of Qiskit, the circuit could not be terminated whenever the ancilla qubit used was measured to be 1. Thus a conditional quantum circuit was created where during each time step the ancilla qubits were measured and the values assigned to classical bit registers. If any of the measurements during the time step were not 0 then the classical bits were all set to 1 and the rest of the time steps no evolution circuits were implemented. This ``marker state" was subsequently used to discard runs after all the Trotter steps were implemented.}

\REV{Results for the depolarizing and Quantinuum H2 noise models indicate relative closeness to those done with the noiseless simulation. This is further corroborated by the measured success probabilities for those three simulations being very close to one another along with the measured $\hat{N}$ value increasing only slightly. Due to the increased noise for the simulated IBM quantum processor, it had a much lower measured success probability per time step and also demonstrated "fermionic occupation smearing" where the population on each of the qubits spread out more evenly among the lattice sites along with a significant noise generated increase of $\hat{N}$ of the lattice as shown in \cref{fig:sucprob}. It should be noted that the $\hat{N}$ could serve as another method of performing further post-selection by removing shots that don't have approximately the same initial value of $\hat{N}$ as the initial state.}

\begin{figure}
    \centering
    \includegraphics[width=0.48\linewidth]{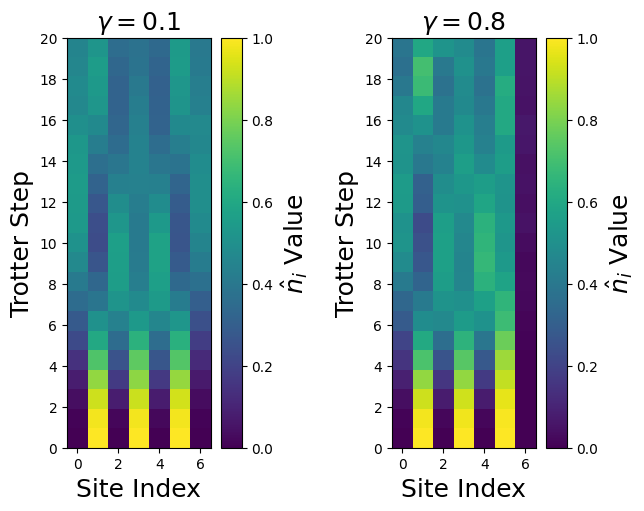}
    \caption{Site fermionic occupation time dynamics for the interacting Hatano-Nelson model in \eqref{eq:interacting_hatano_nelson} $(J = 1.0, \gamma = 0.1\ \&\ 0.8, V_{i,i+1} = 2.0, \tau = 0.1)$ with only site indices 5 \& 6 having the non-Hermitian $\gamma$ dependent term applied. Quantum circuit simulations of the time dynamics were performed noiselessly.}
    \label{singlesite_pop}
\end{figure}

\begin{figure}
    \centering
    \includegraphics[width=0.44\linewidth]{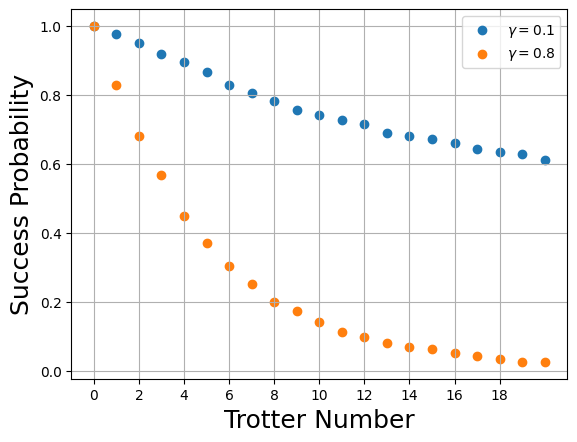}
    \caption{ Success probability for the algorithm implementing a multi time step evolution for the interacting HN model $(J = 1.0, \gamma = 0.1 \ \&\  0.8, V_{i,i+1} = 2.0, \tau = 0.1)$ with only site indices 5 \& 6 having the non-Hermitian term $\gamma$ applied. Success probability is computed the same way as in \cref{fig:sucprob} }
    \label{singlesite_sucprob}
\end{figure}

\REV{As a final experiment, we perform noiseless simulations of the algorithm when only a single link of the 1D chain has a non-Hermitian term. This is done to investigate how the success probability changes given a reduction in the non-Hermitianness of the Hamiltonian of interest. Simulations are performed with the same lattice size, time step, and interaction strength as those done in \cref{fig:skineffect}. The resulting population time dynamics and success probability are given in \cref{singlesite_pop} and \cref{singlesite_sucprob} respectively. The two significant points to note are the decrease in the success probability as $\gamma$ increases and the change in the reflection of any particle movement away from the sites where non-Hermitian terms are applied. Since the non-Hermitian term acting on a link causes a particle populated on that link to move left, applying the non-Hermitian term to the rightmost link (sites 5 \& 6) essentially causes any particles to be reflected away from those sites and thus reduces the effective lattice size. This is evident by the right plot in \cref{singlesite_pop} having near no population at site index 6. Given that the success probability is considerably larger with only single non-Hermitian term, this is indicative of the potential of our algorithm for elucidating longer time behavior of general systems governed by dynamics following \cref{eq:ode_general} when there are fewer $V_j$ terms or their characteristic strength is lower. }

\REV{There are many interesting further avenues of investigation to explore for implementing this algorithm, most notably executing on actual quantum hardware. Given the performance of the algorithm on a simulated IBM Torino device, devices with better 2Q fidelity such as the Quantinuum QCs would be preferable. Additionally, further circuit optimization, error mitigation, error detection, and error correction techniques could be implemented to improve performance. It is likely that applying these techniques to the ancilla qubit used for measurement would have an outsized impact on the performance of the algorithm by reducing false non-zero measurements on the non-Hermitian associated evolution terms.}

\REV{For the low and no noise simulations (noiseless, depolarizing, Quantinuum), the exponential fit $p_s = \rho^s$ (where $p_s$ is the success probability up to step $s$ and $\rho$ is the estimated success probability for each step) for the success probability diverged from the measured success probability at the 4th Trotter step. This was not the case for the IBM simulation as the fitted exponential matched the observed success probability quite well. Further inquiry into this result would help elucidate the behavior of the algorithm in the longer time limit.}

\REV{Investigations into the dependence of the measured success probability on the non-Hermitian coefficients $\gamma$, the lattice size $N$, and the noise added to the simulation would be desired. Initial experiments were run for smaller models and it was observed there was a negative correlation between all of these variables and the fitted probability value for the time-dependent success probability. It may be possible to develop a phenomenological model for the success probability in terms of $\gamma$, $N$, and some characteristic noise coefficient such as the depolarizing channel probability $p$. This would give a better understanding of the limit of problems that can be addressed given current hardware shot count limits.}

\REV{Once a better understanding of the algorithm's performance on hardware and improved strategies for reducing errors are obtained, it will serve as a useful algorithmic primitive for accurately simulating properties of other non-Hermitian Hamiltonians. Such examples include simulating the non-Hermitian Su–Schrieffer–Heeger model and its topological properties. Topological invariants are important properties of non-Hermitian Hamiltonians and include quantities such as winding and Chern numbers, Berry phase, etc. Measurements of these properties have previously been performed for Hermitian systems on quantum computers \cite{xiao_robust_2023}. The numerical simulations performed above indicate that this algorithm will have significant utility for elucidating the dynamics of non-Hermitian Hamiltonians on near-term quantum processors.}

\section{Conclusion and Discussion}

In this work we propose a quantum algorithm for solving linear ODEs. Our algorithm has the desirable features that it uses a single ancilla qubit and is locality preserving, i.e., and when the coefficient matrix $A$ is $k$-local, our algorithm only need to implement the time evolution of $(k+1)$-local Hamiltonians. These features make our algorithm suitable for implementation on near-term and early fault-tolerant devices. Interestingly, our algorithm has a close connection to Lindbladian dynamics.

The locality-preserving property of our algorithm makes it especially useful for simulating non-Hermitian Hamiltonians, and we demonstrate its utility by discussing the simulation of the interacting Hatano-Nelson model in detail.

Although our algorithm, being a first-order method, does not achieve the optimal runtime \cite{AnChildsLin2023quantum}, it is possible that advanced post-processing techniques can make its runtime match the state-of-the-art algorithms when estimating observable expectation values.
Specifically, using data generated from finite step sizes, one can extrapolate towards the zero-step-size limit. This approach has been explored for the Trotter decomposition in the context of Hamiltonian simulation with great success \cite{RendonWatkinsWiebe2024,WatsonWatkins2024,Watson2024}, and much of the analysis can also be applied to the present setting.

Our algorithm may be an ideal candidate for demonstrating the 
improving utility and capability of near-term quantum devices. One such example is mid-circuit measurement (MCM), the ability to selectively measure a subset of qubits on a quantum computer without decohering any other qubits. MCM is an important component in our algorithm and can also be used in other algorithmic subroutines such as rapid preparation of states with long-range entanglement \cite{TantivasadakarnetAlPRXQuantum2023}, measurement based quantum computing \cite{LanyonetAlPRL2013}, and quantum error correction \cite{Chiaverini2004Nature, AndersonetAlPRX2021}.
It has been implemented in many of the major qubit modalities such as neutral atoms \cite{NorciaetAl2023PRX}, trapped ions \cite{Chiaverini2004Nature, AndersonetAlPRX2021}, superconducting qubits \cite{GambettaPRL2021}, etc. For future work, it is of interest to implement our algorithm on real quantum devices to probe physical phenomena such as the non-Hermitian skin effect, which is otherwise difficult to demonstrate in experiments. 

\REV{Whether solving differential equations, aside from Hamiltonian simulation, on a quantum computer can lead to significant quantum advantage has been an active area of research. Recently examples of differential equations describing classical dynamics that are \textsf{BQP}-complete have been found \cite{BabbushBerryKothariSommaWiebe2023,BravyiMansonSawkoEtAl2025quantum}. This means these differential equations can be solved efficiently on a quantum computer, while should not be efficiently solvable using any classical algorithm, as otherwise the efficient classical algorithm would be able to efficiently simulate any polynomial-time quantum computation. However, these examples all require large-scale fault-tolerant quantum computers, and so far we do not know any example of quantum advantage in differential equations that can run on near-term hardware. If such an example can be found then the techniques in this paper may be helpful for its experimental implementation.}

\REV{This work also provides another perspective for the utility of solving differential equations on quantum computers: instead of achieving a computational advantage over classical computers, it may provide a tool to study interesting physics on current hardware, much like recent experiments realizing time crystals \cite{MiIppolitiEtAl2022time} and non-abelian topological order \cite{IqbalTantivasadakarn2024non}, or the non-Hermitian skin effect studied in~\cref{sec:appl_Hatano-Nelson}. Our techniques offer great flexibility to perform short evolution under non-Hermitian Hamiltonians and for state preparation, making it potentially useful for beyond-classical quantum simulation of interesting physical phenomena.}

\section*{Acknowledgments}
We thank Jungsang Kim and Thomas Barthel for helpful discussions. 
This work is supported by the U.S. Department of Energy, Office of Science, Accelerated Research in Quantum Computing Centers, Quantum Utility through Advanced Computational Quantum Algorithms, grant no. DE-SC0025572, and National Science Foundation via the grant DMS-2347791 and DMS-2438074 (D.F.).

\appendix
\section{Application to the Convection-Diffusion Equation}
\label{sec:convection_diffusion}

\REV{In this section we discuss the application of our algorithm to the convection-diffusion equation. Specifically we focus on how to decompose the coefficient matrix $A$ into the form given by \eqref{eq:LL_decompose}. In other words, we want to efficiently find $L_j$ and $H$ such that
\[
A = -iH - \sum_{j=1}^J L_j^\dag L_j.
\]
}

\REV{
We consider the convection-diffusion equation:
\begin{equation}
\label{eq:convection_diffusion}
    \partial_t c(x,t) = Ac(x,t) := -\nabla\cdot (\mathbf{v}(x)c(x,t)) + \frac{1}{2}\Delta c(x,t),
\end{equation}
where $\mathbf{v}(x)$ is a velocity field. 
For simplicity we consider this PDE on a $D$-dimensional torus.
}

\REV{
One can decompose the generator $A$ into a self-adjoint (corresponding to Hermitian) part and an anti-self-adjoint (corresponding to anti-Hermitian) part, $A=A_+ + A_-$:
\[
A_+ c = -\frac{1}{2}(\nabla\cdot \mathbf{v}) c + \frac{1}{2}\Delta c,\quad A_- c = -\frac{1}{2}(\nabla\cdot(\mathbf{v}c) + \mathbf{v}\cdot \nabla c).
\]
The anti-self-adjoint part can be simulated as a Hermitian Hamiltonian by setting $H=iA_-$, so we will focus on the self-adjoint part $A_+$. This term consists of two parts: $-\frac{1}{2}(\nabla\cdot\mathbf{v})c$ and the Laplacian $\frac{1}{2}\Delta c(x,t)$. $-\frac{1}{2}(\nabla\cdot\mathbf{v})$ is diagonal in the grid-point basis (it is simply point-wise application with a function), so one can simply take 
\[
h(x) = \sqrt{\nabla\cdot\mathbf{v}(x)-\min_{y}\nabla\cdot\mathbf{v}(y)},
\]
and define 
\[
L_0 c(x,t) = \frac{1}{\sqrt{2}}h(x)c(x,t).
\]
For the Laplacian, we have
\[
\frac{1}{2}\Delta  = -\sum_{j=1}^D \left(\frac{i}{2}\partial_{x_j}\right)^2.
\]
Therefore we can let
$
L_j = \frac{i}{2}\partial_{x_j},
$
for $j=1,2,\cdots,D$. From the above we can see that
\[
A_+ = -\sum_{j=0}^D L_j^2 - \min_{y}\nabla\cdot\mathbf{v}(y),
\]
and therefore
\[
A = A_- -\sum_{j=0}^D L_j^2 - \min_{y}\nabla\cdot\mathbf{v}(y)
\]
gives us the desired decomposition. In this example all jump operators $L_j$ are self-adjoint, but this may not always be true in the general case.
}

\REV{
To simulate \eqref{eq:convection_diffusion} on a quantum computer, one would need to apply spatial discretization to turn it into an ODE. There are many ways to accomplish this. In the simplest case, one can use finite-difference to discretize $c(x,t)$ into a function over grid points. Another very commonly used method is the spectral method where discretization can be done in the Fourier basis. In either case the jump operators are easy to implement because they are either diagonal in the grid-point basis ($L_0$) or the Fourier basis ($L_j$, $j\geq 1$), and these two bases are related to each other through the Fourier transform, which can be implemented efficiently on a quantum computer.
}

\bibliographystyle{unsrt}
\bibliography{ref_all}
\end{document}